\documentclass{article}

\usepackage[dblblindworkshop, final]{neurips_2026}

\usepackage[utf8]{inputenc} 
\usepackage[T1]{fontenc}    
\usepackage{hyperref}       
\usepackage{url}            
\usepackage{booktabs}       
\usepackage{amsfonts}       
\usepackage{nicefrac}       
\usepackage{microtype}      
\usepackage{xcolor}         
\usepackage{graphicx}
\usepackage{amsmath}

\title{Similarity Pairing with Energy Mover's Distance for Self-Supervised Pre-Training at the LHC}

\author{%
  Ho Fung Tsoi, Dylan Sheldon Rankin \\
  Department of Physics and Astronomy \\
  University of Pennsylvania, USA \\
  \texttt{\{hftsoi,dsrankin\}@sas.upenn.edu} \\
}

\begin{document}

\maketitle

\begin{abstract}
Many self-supervised methods for training foundation models at the Large Hadron Collider (LHC) rely on data augmentations to encourage the model to embed events into a representation space invariant to certain physical or detector symmetries.
A common challenge arises from the large freedom in choosing a proper set of augmentations on which downstream performance depends.
The implementation of augmentations involves either modifying existing events, potentially breaking the event fidelity, or simulating more event variants, which is computationally intensive.
In this work, we present a data-driven method of pairing events by their similarity via the energy mover's distance (EMD), which measures how similar two events are in terms of the work required to transform one into the other.
With this approach, distinct events are sampled and matched by their similarity to serve as views for learning invariance, keeping the physics content of each event intact without handcrafted distortions.
We demonstrate this augmentation-free pairing method by pre-training on QCD jets via self-distillation and show that it can yield semantic jet embeddings with downstream discrimination power comparable to or better than an augmentation-based baseline.
\end{abstract}

\section{Introduction}

Self-supervised learning (SSL) has emerged as a common underlying technique for foundation model training.
The typical workflow starts with pre-training a model on a large amount of unlabeled data without a specific downstream task to learn a generic representation space, then fine-tuning the model on labeled data to perform various downstream supervised tasks.
SSL techniques include contrastive learning~\citep{9157636,chen2020simple}, embedding regularization~\citep{bardes2022vicreg}, and self-distillation~\citep{caron2021emerging,zhou2021ibot,oquab2023dinov2}.
Most methods begin by creating different views from each example through data augmentations, encouraging the model to learn invariance up to the handcrafted distortions or symmetries, so the representation space tends to encode high-level semantics and ignore low-level details.

The SSL paradigm has been adopted in the high-energy physics (HEP) field, ranging from improving model performance against mismodeling in simulation to training general-purpose foundation models~\citep{Golling:2024abg,10.21468/SciPostPhys.12.6.188,Harris:2024sra,SciPostPhys.18.5.150,hao2025rino} (see \citep{hallin2025foundation} for a recent review).
Unlike in computer vision, where distortions of images in various forms can be freely made without many constraints, the augmentations for HEP data modeling require respecting the underlying physical laws and detector constraints, and handcrafted augmentations can easily break physical symmetries in the data and make the model unreliable if they are not carefully designed.
In this work, we present a data-driven way of forming ``augmented'' pairs without handcrafted augmentations, by pairing similar events with the energy mover's distance (EMD), so the underlying physics in events is unaltered.
We pre-train a jet model with an augmentation-based SSL method, but replace the augmentation with the similarity pairing and show that the embedding yields meaningful clustering structure.
We perform downstream probing with an anomaly detection task and show that it can yield performance comparable to or better than an augmentation-based baseline under the same SSL method, demonstrating its potential for integration into future pre-training workflows.

\section{Method}
Inspired by MACK~\citep{SciPostPhys.18.5.150}, which proposed to use EMD to pair real data events and simulated events in SSL training to reduce the effects of systematic uncertainties on model performance caused by mismodeling in simulation, we adapt the idea to generic embedding learning, making it potentially useful for foundation model training.

\textbf{Data-driven EMD pairing as ``augmentations''.}
In a given augmentation-based SSL pre-training method, we remove the augmentation step and replace it with the data-driven similarity pairing.
Instead of creating two distorted copies of an event, a training pair is formed from two distinct events that are close in the EMD, which is an optimal-transport metric measuring the work required to transform the energy distribution from one event to another~\citep{Komiske:2019fks}.
Thus, a small enough EMD of a pair means that their energy distributions are geometrically similar, which can be viewed as ``the same event up to fluctuations''.
Here the fluctuations are not manually introduced through augmentations but are genuine differences between two real events, whose physics content is kept intact.
In this way, when the model is trained on real collision events, the underlying physics is kept unaltered, with no risk of breaking physical laws or detector constraints as could result from augmentations.
We use the Wasserstein package~\citep{10.1145/2070781.2024192,Komiske:2019fks,Komiske:2020qhg} to compute the pairwise EMD metric.

\textbf{Self-supervised pre-training.}
The pairing method we propose is agnostic to the self-supervised method as long as it requires creating augmented view pairs for invariance learning in the pre-training.
In this work, we use the jBOT~\citep{SciPostPhys.21.3.053} method, which is based on self-distillation following the iBOT~\citep{zhou2021ibot} framework from computer vision, to perform pre-training on jet data.
The pre-training starts by pairing two distinct jets via the EMD, after which each pair is fed to a teacher-student encoder architecture.
The student processes masked versions of jets and then tries to predict the representations produced by the teacher, at both the particle and jet levels.
The main objective is to minimize the cross-entropy between the teacher and student output distributions.
We largely follow the pre-training setup of jBOT except for the formation of training pairs and refer the reader to \citep{SciPostPhys.21.3.053} for the implementation details.

\section{Experiments}

\textbf{Dataset.}
We use the JetNet dataset~\citep{Kansal:2021cqp,kansal_2022_6975118}, which contains jet samples simulated from proton-proton collisions at 13 TeV.
The samples include five jet classes: light quark ($q$), gluon ($g$), $W$ boson, $Z$ boson, and top quark ($t$).
The jets have transverse momentum $p_{\text{T}}$ around 1 TeV, clustered using the anti-$k_{\text{T}}$ algorithm~\citep{Cacciari:2008gp} with distance parameter 0.8.
Each jet is represented by up to 30 particles of highest $p_{\text{T}}$, each with $p_{\text{T}}$ relative to the jet $p_{\text{T}}$, and pseudorapidity $\eta$ and azimuthal angle $\phi$ relative to the jet axis.

\begin{figure}[!t]
    \centering
    \includegraphics[width=0.75\textwidth]{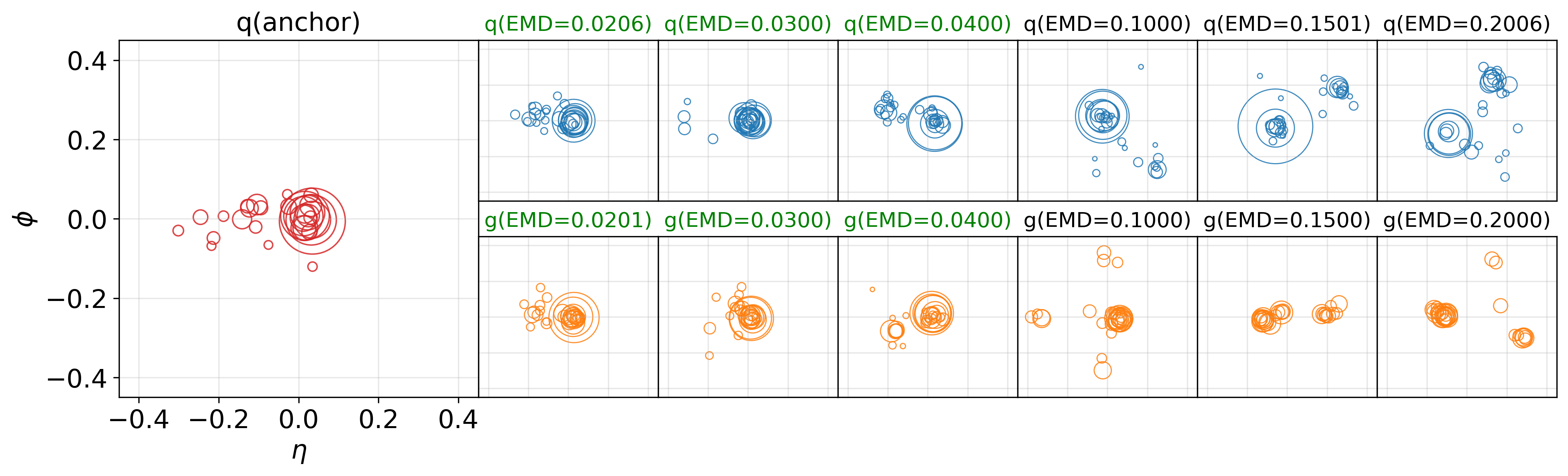}
    \caption{An anchor $q$ jet (left), and example jets sampled at different EMD values from the $q$ class (upper) and $g$ class (lower), respectively. Sampled jets within the pairing window EMD $\in (0.01,0.05)$ are indicated by green titles. In the drawn jet plane, each particle is represented by a circle whose center coincides with the particle coordinate, and the circle size is proportional to the particle $p_{\text{T}}$.}
    \label{fig:emd-examples}
\end{figure}

\begin{figure}[!t]
    \centering
    \includegraphics[width=0.2\textwidth]{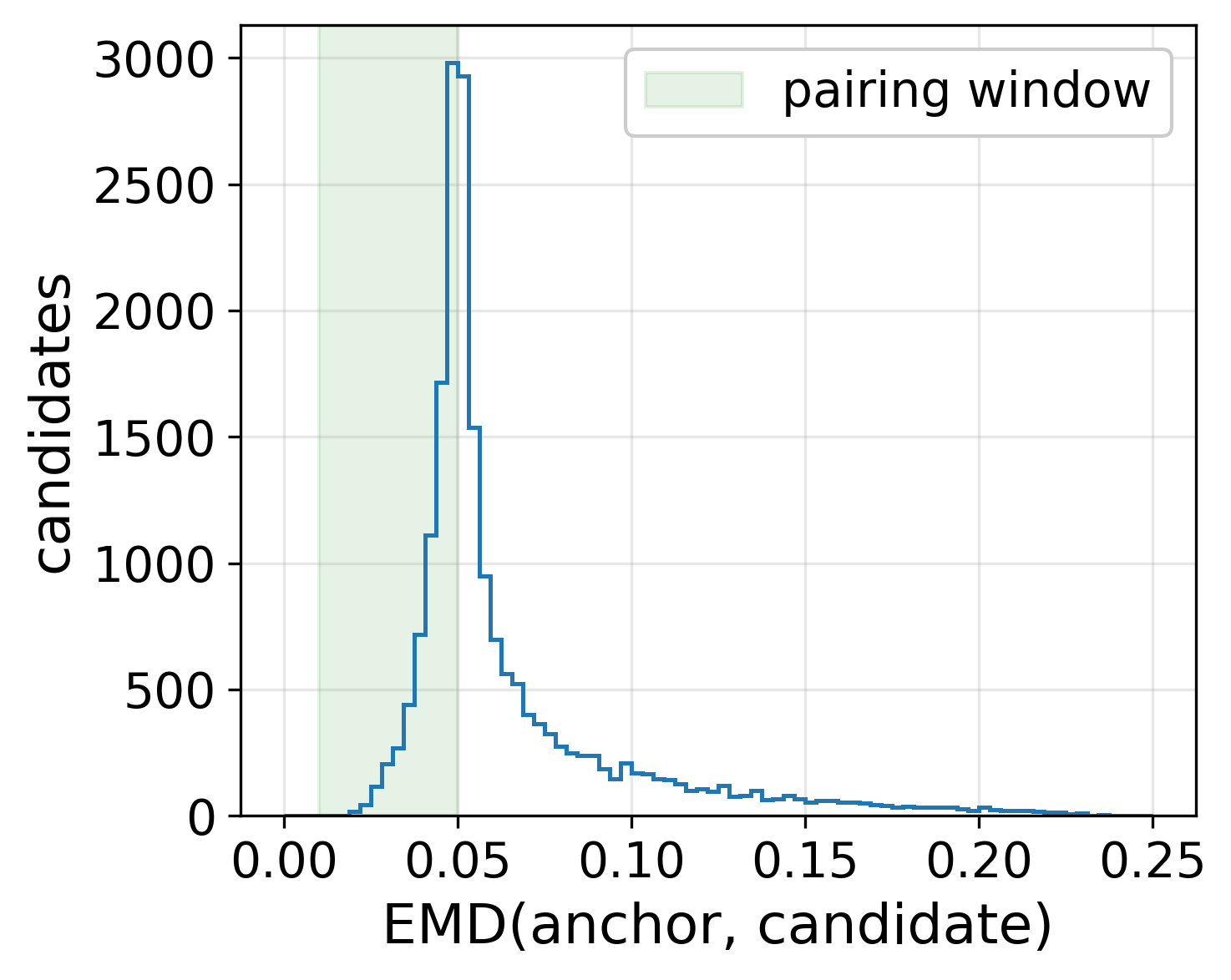}
    \caption{EMD distribution of the anchor jet in Fig.~\ref{fig:emd-examples} with other QCD jets in the sample. The green window indicates the EMD range used for pairing in the pre-training.}
    \label{fig:emd-window}
\end{figure}

\textbf{Pairing schemes and pre-training.}
We pre-train on QCD jets only ($q$ and $g$) and then directly probe the embedding for downstream anomaly detection without fine-tuning.
There are 880k jets in the dataset and we take 75\% of the $q$ and $g$ jets for the pre-training; the rest of the jets, with an equal number per class, are used for the test set.
We explore three pairing modes: (1) same-class pairing where $q$ pairs to $q$ only and $g$ pairs to $g$ only, (2) different-class pairing where $q$ pairs to $g$ only and $g$ pairs to $q$ only, and (3) random pairing where there is no constraint.
The EMD is computed with normalized $p_{\text{T}}$ and angular distances scaled by the jet radius $R=0.8$.
Fig.~\ref{fig:emd-examples} shows a $q$ jet as an anchor, and six example jets from each of the $q$ and $g$ classes, from small to large EMD values for each pair, showing that the similarity in the distributions is higher for lower EMD.
Fig.~\ref{fig:emd-window} shows the EMD distribution across the whole training sample for the same anchor jet.
We use the EMD window (0.01, 0.05) for the pairing in pre-training.
Typically for a given jet, roughly half of the candidates fall within the window.
Therefore, the pairing replaces the large degree-of-freedom augmentation space with a single window choice, and we find the performance stable under moderate variations in the window definition as long as it is not too narrow or too wide.
We embed each particle into a 32-dimensional vector and pre-train a transformer~\citep{vaswani2017attention} encoder with three layers, each with four attention heads and a feedforward layer with 128 dimensions, with GELU activation~\citep{hendrycks2016gelu} and dropout~\citep{JMLR:v15:srivastava14a} at 0.2.
A [CLS] token prepended to the list of particles summarizes each jet's semantics and serves as the overall jet embedding for downstream probing.
The model is implemented with Keras~\citep{chollet2015keras}, trained with AdamW~\citep{loshchilov2018decoupled} for 100 epochs at learning rate $2\times 10^{-3}$ and weight decay $10^{-4}$, with a cosine decay schedule~\citep{loshchilov2017sgdr} and batch size 1024.

\begin{figure}[!t]
    \centering
    \includegraphics[width=0.2\textwidth]{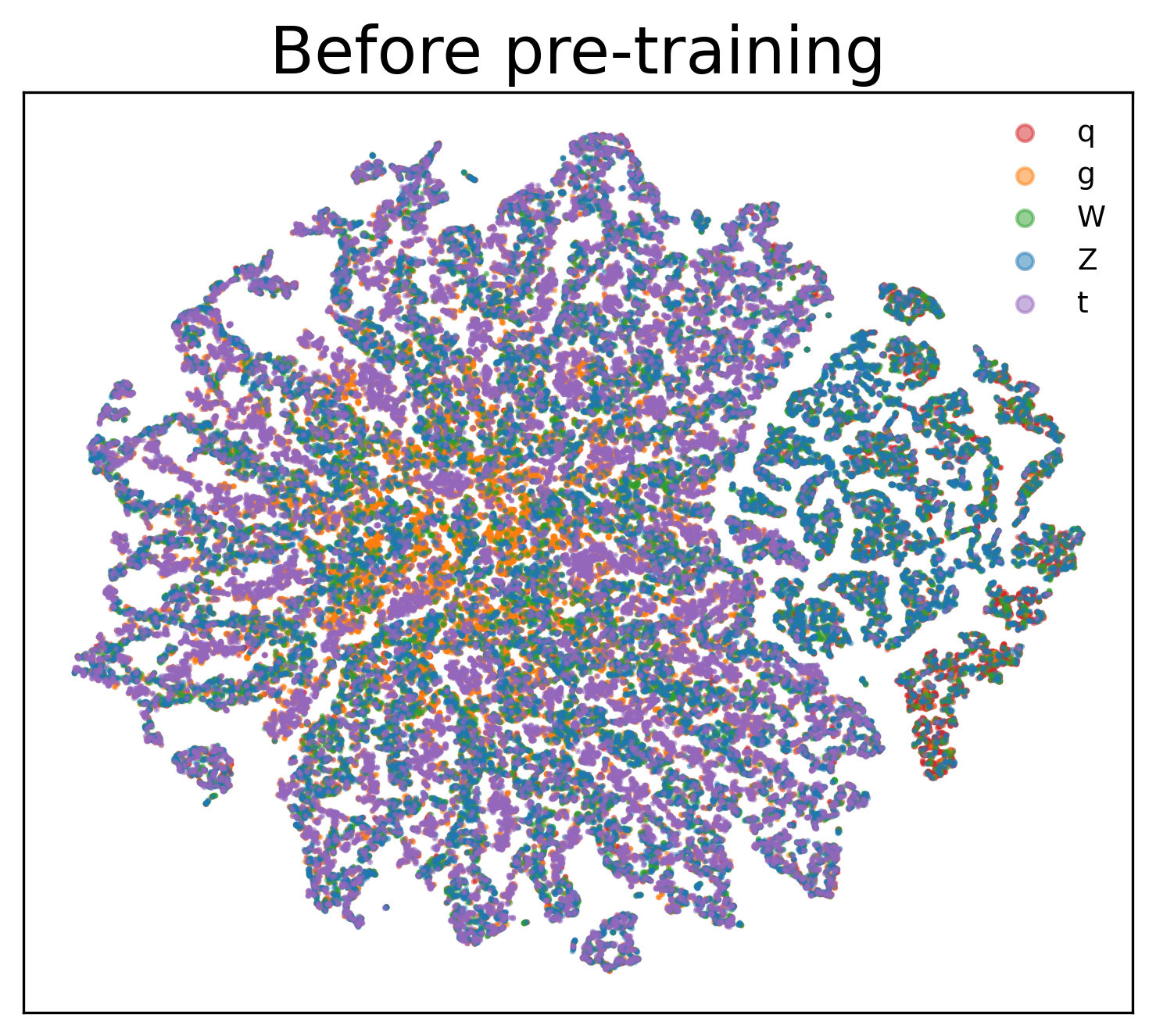}
    \includegraphics[width=0.2\textwidth]{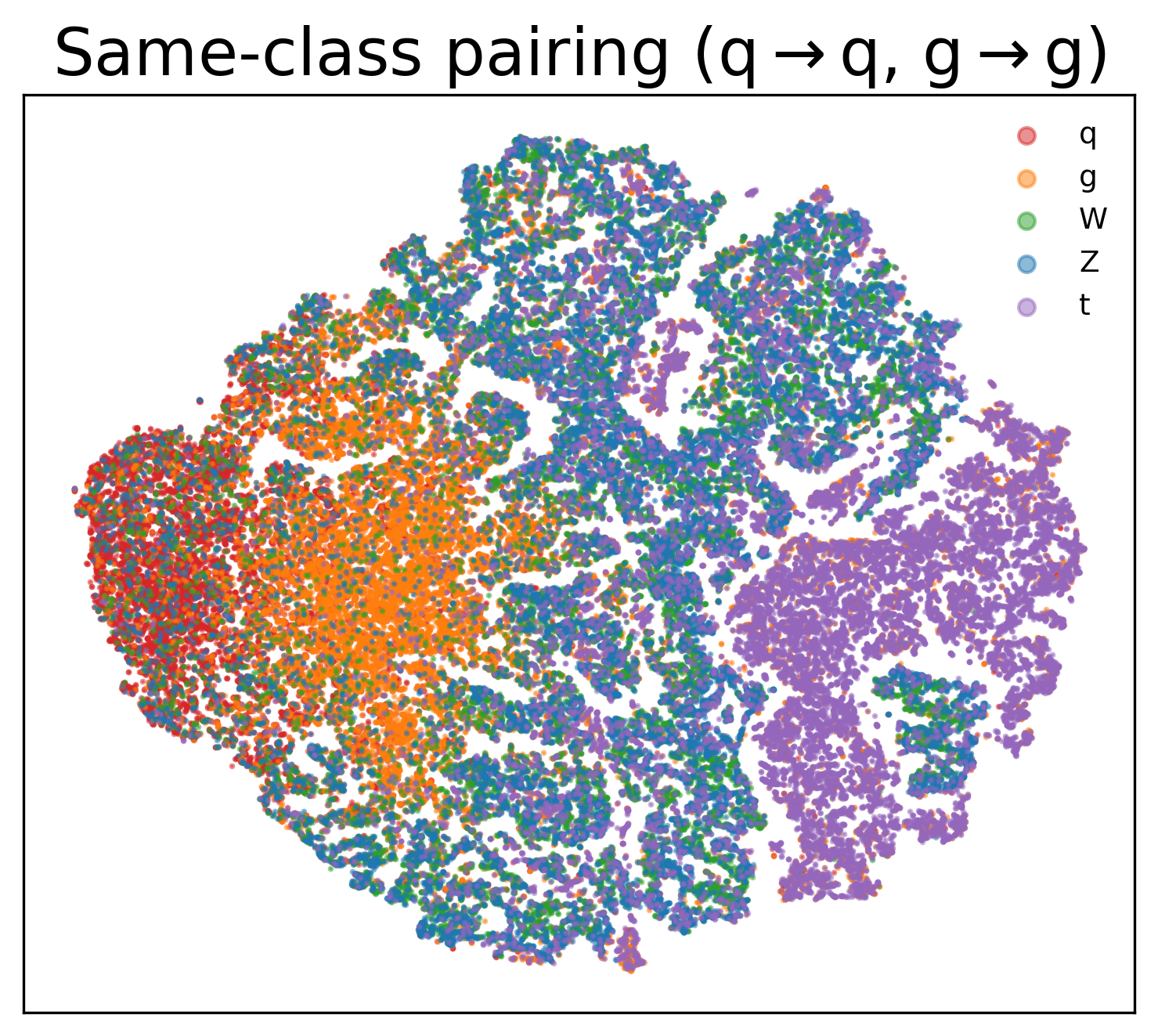}
    \includegraphics[width=0.2\textwidth]{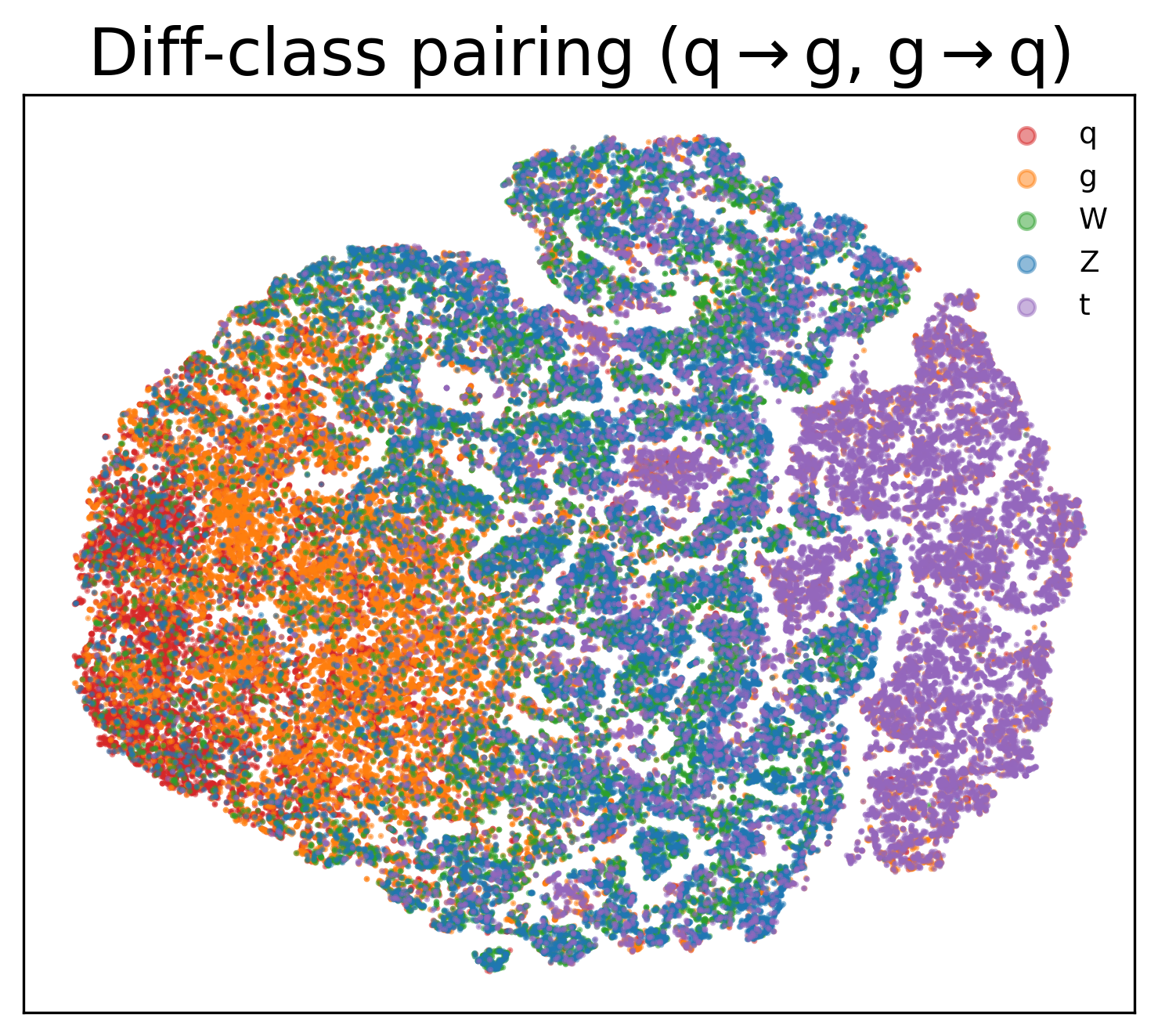}
    \includegraphics[width=0.2\textwidth]{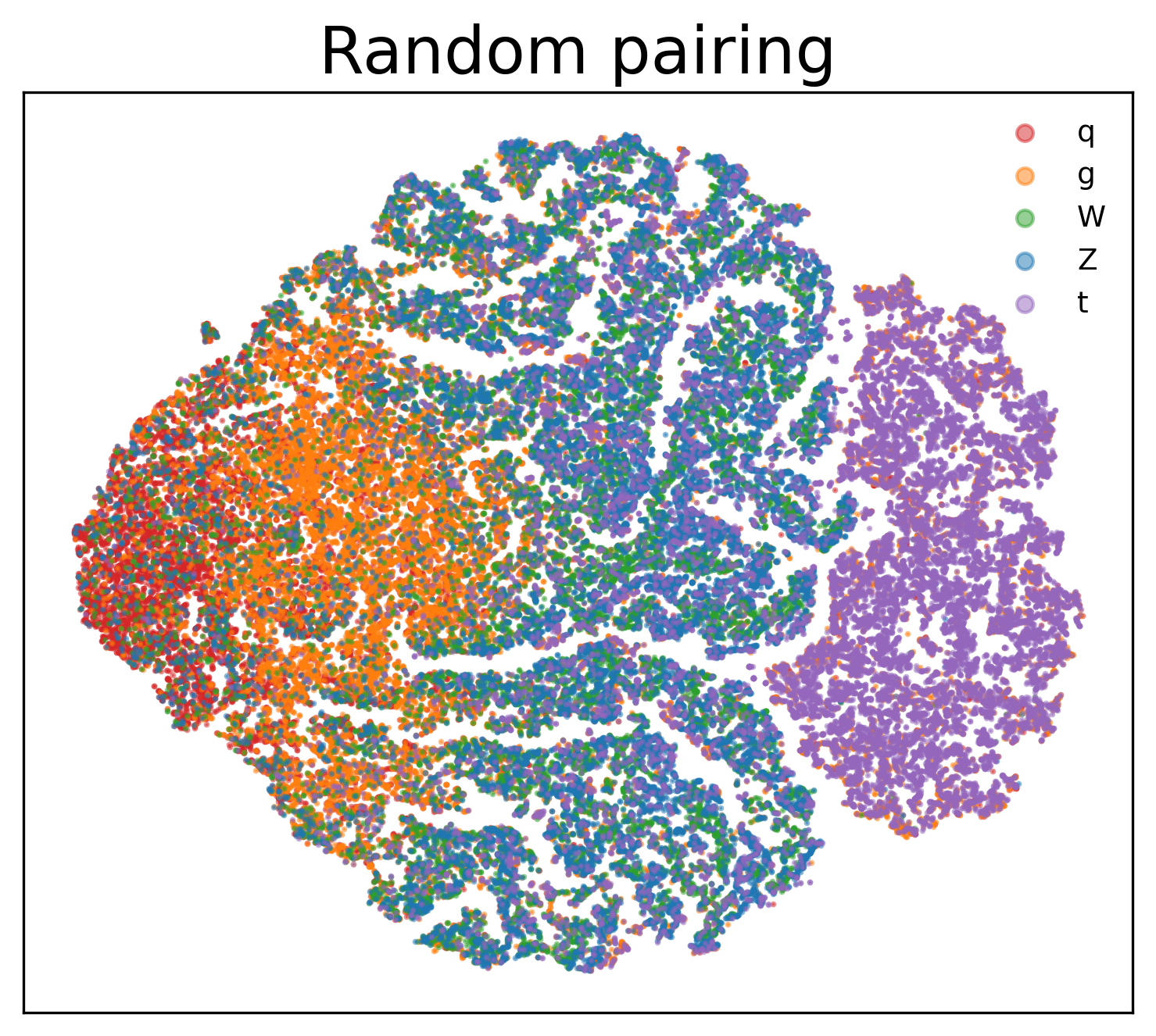}
    \caption{t-SNE projections showing clustering in the pre-trained embedding: $q$ (red), $g$ (orange), $W$ (green), $Z$ (blue), and $t$ (purple). Training is done on $q$ and $g$ jets only, with three different pairing modes separately, without using labels except for the pairing stage in the same-class and diff-class modes. The leftmost is the randomly initialized model before pre-training, included as a reference.}
    \label{fig:tsne}
\end{figure}

\textbf{Clustering structures in the pre-trained embedding.}
Fig.~\ref{fig:tsne} visualizes the pre-trained embeddings in 2D using t-SNE~\citep{JMLR:v9:vandermaaten08a}.
A clear clustering structure appears with class separation, not just between $q$ and $g$ jets but also among the other classes not used in the pre-training.
The $q$ and $g$ jets appear more localized and densely packed than in the original jBOT embedding under a comparable setup, as the pairing pulls QCD jets closer together.
The clustering structure does not differ much among the three pairing modes.

\begin{figure}[!t]
    \centering
    \includegraphics[width=0.22\textwidth]{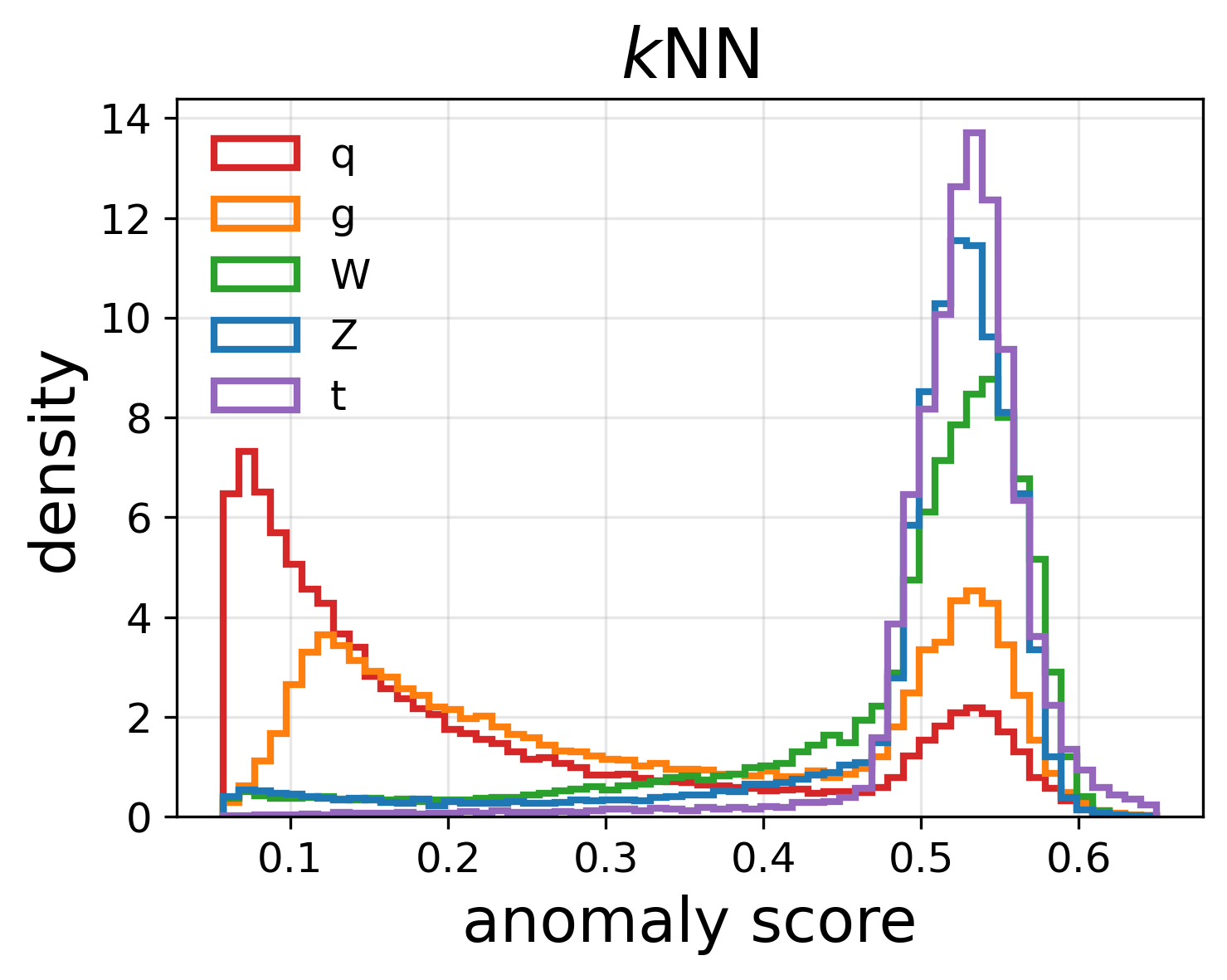}
    \includegraphics[width=0.22\textwidth]{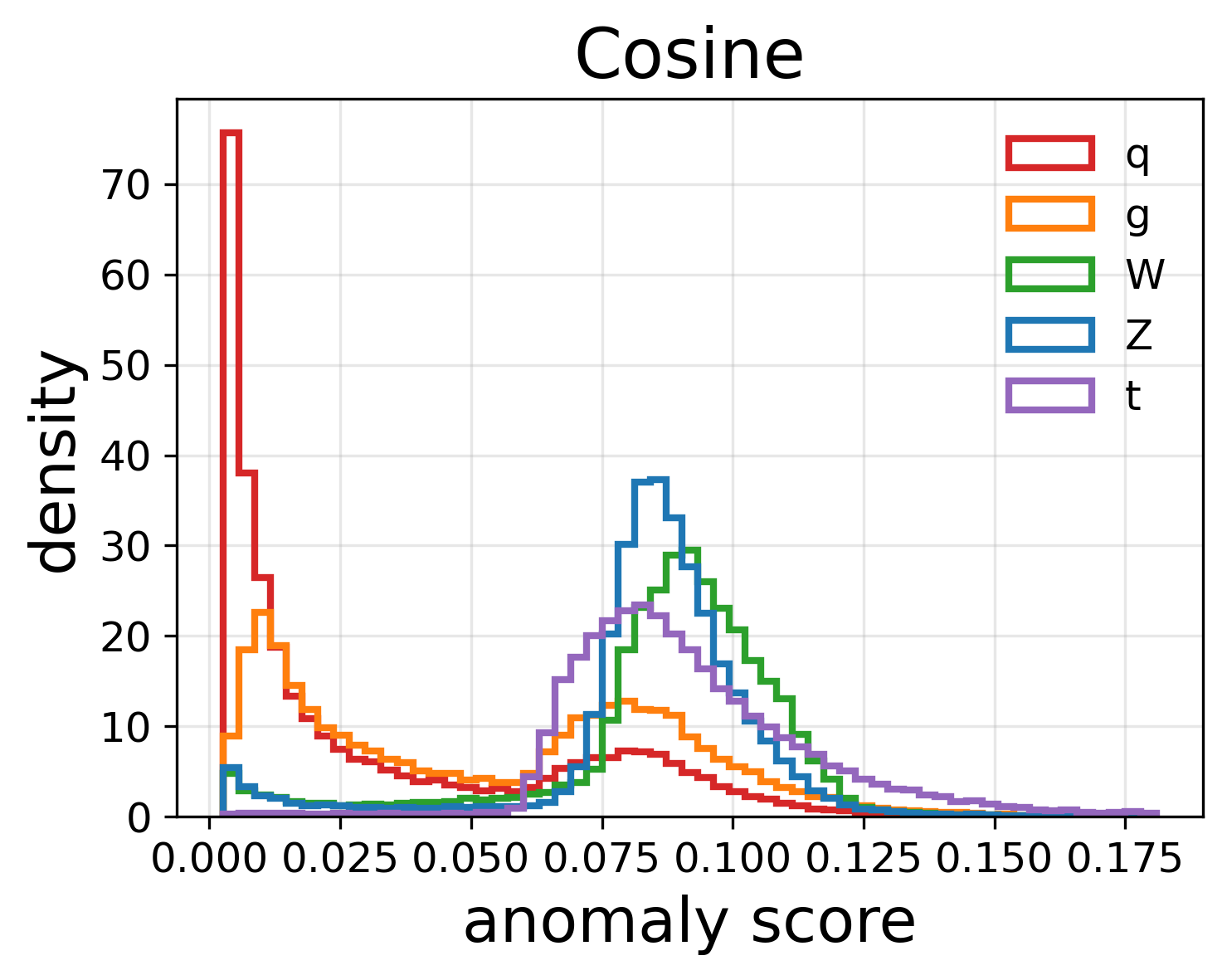}
    \includegraphics[width=0.22\textwidth]{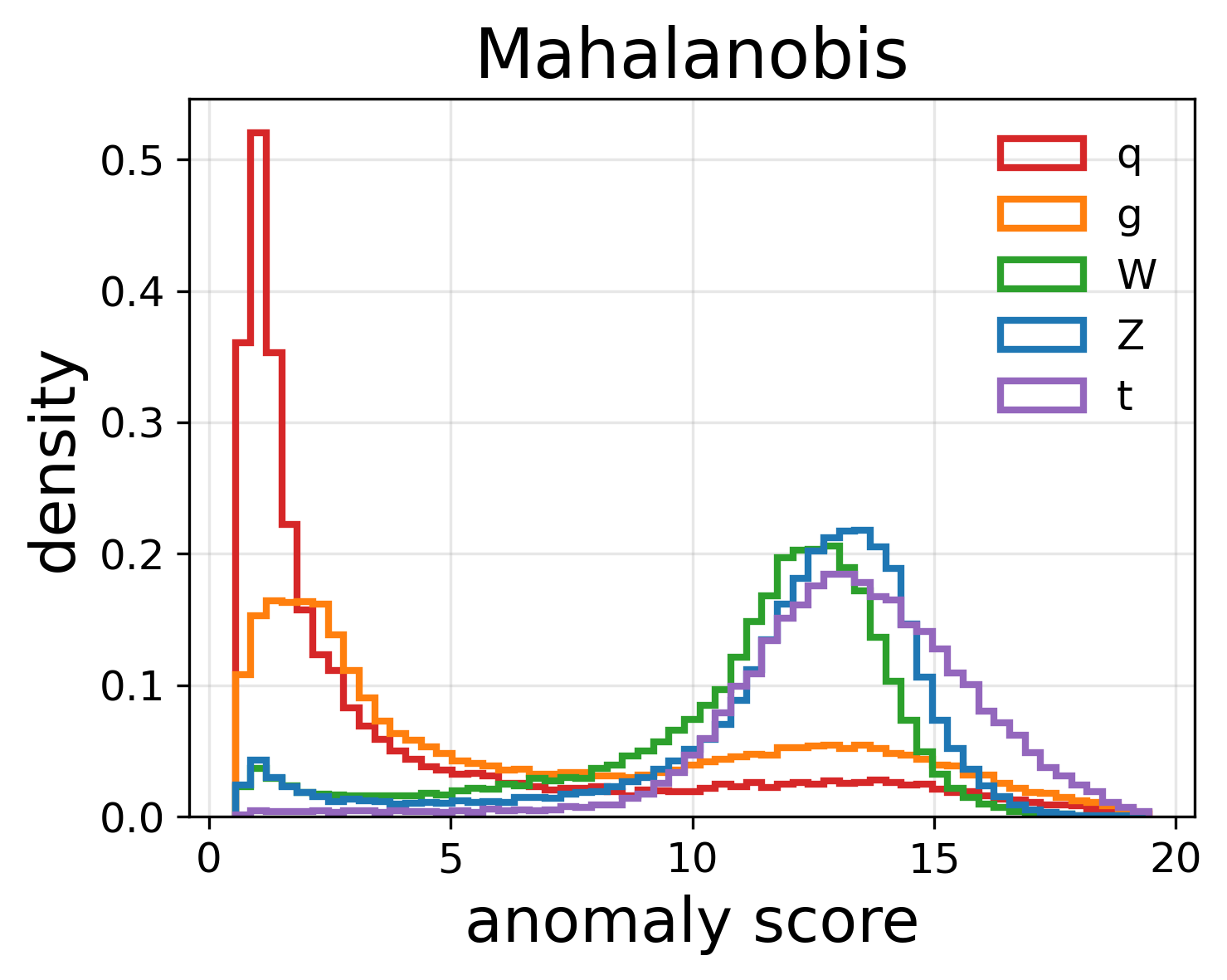}
    \caption{Anomaly score distributions of the three metrics for the final random-pairing model, for the QCD background ($q$ and $g$) and the individual signal classes ($W$, $Z$, and $t$).}
    \label{fig:score}
\end{figure}

\begin{figure}[!t]
    \centering
    \includegraphics[width=0.22\textwidth]{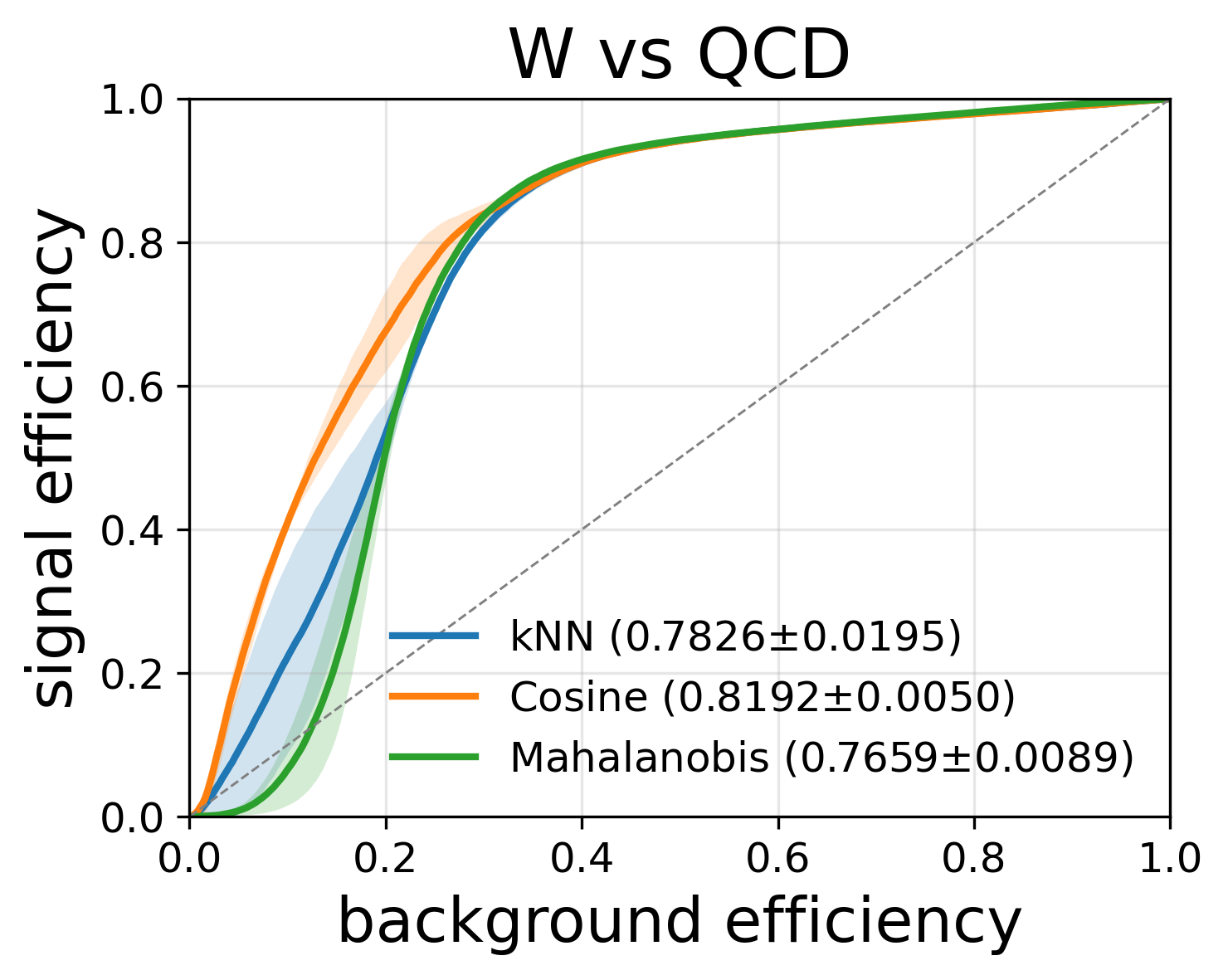}
    \includegraphics[width=0.22\textwidth]{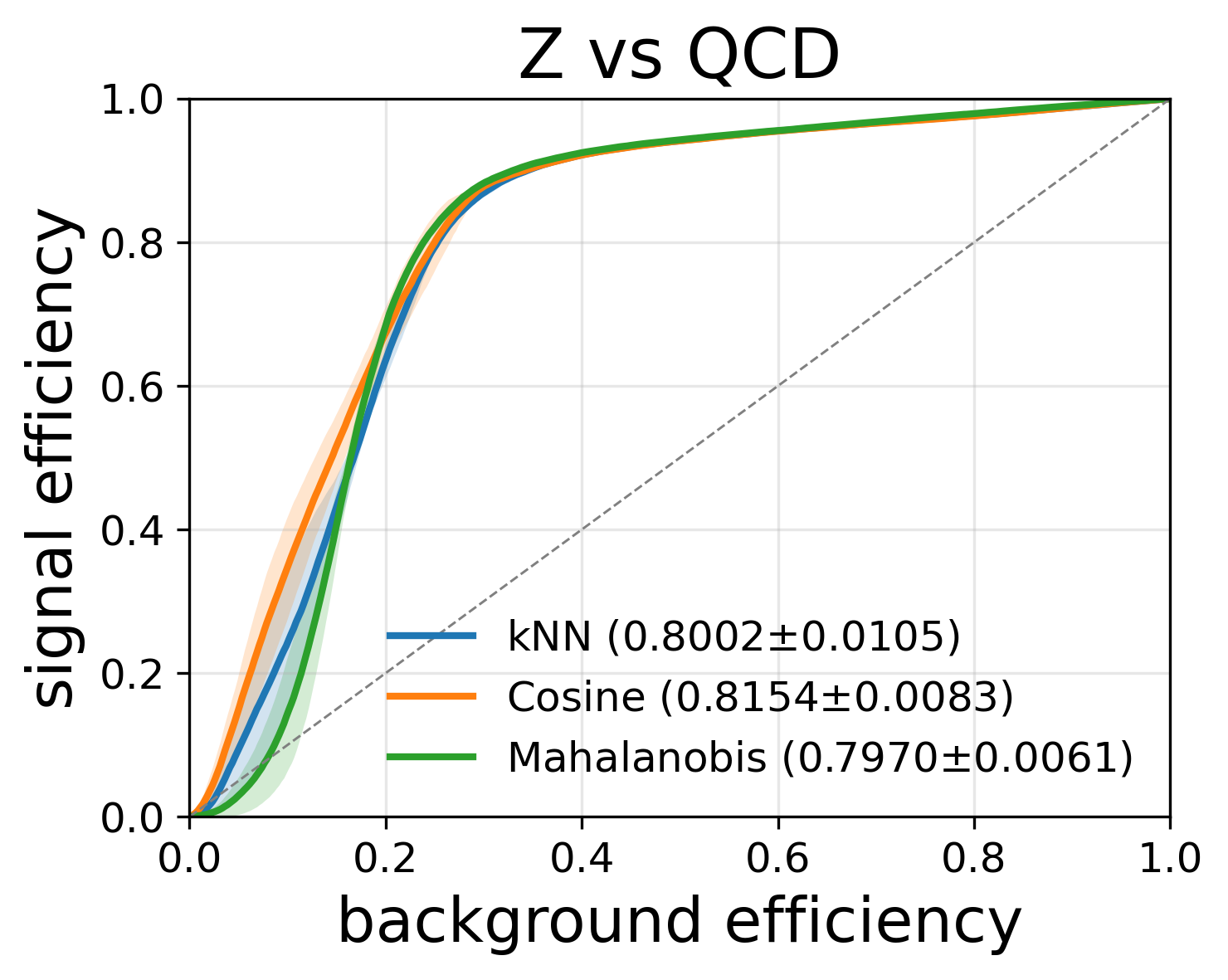}
    \includegraphics[width=0.22\textwidth]{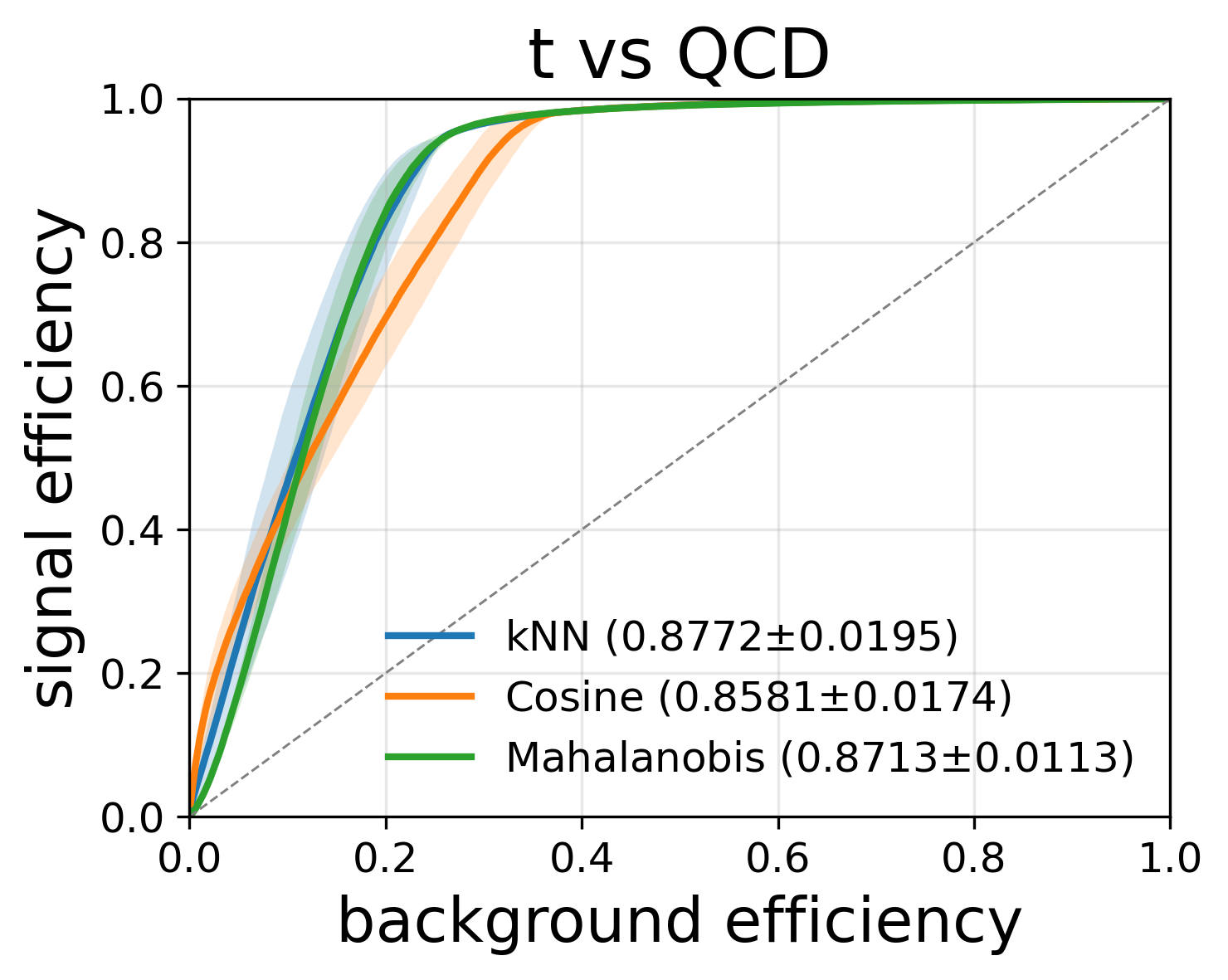}
    \includegraphics[width=0.22\textwidth]{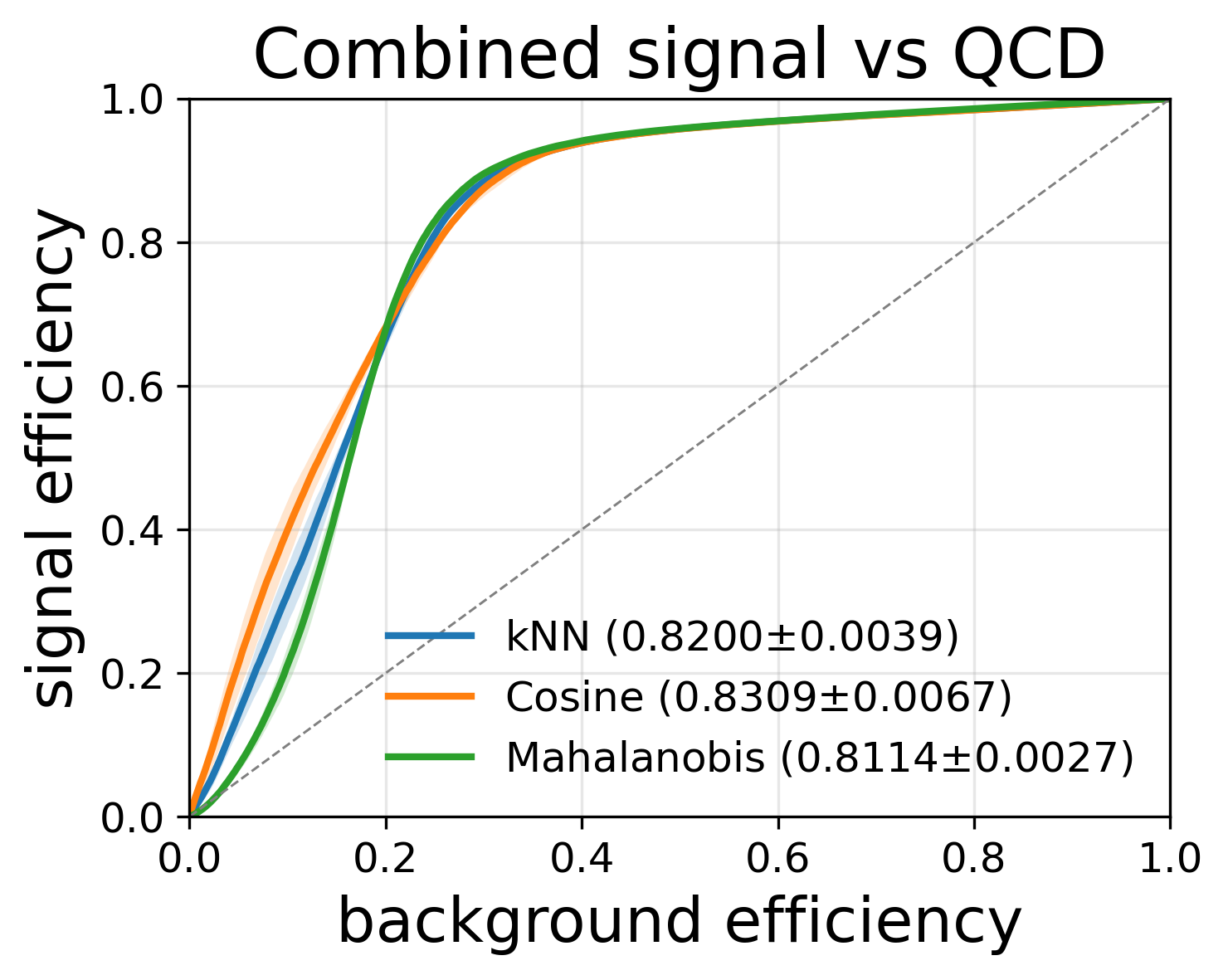}
    \caption{ROC curves for individual signals and combined signal. The curves and the bands represent the mean and standard deviation over three independent pre-trainings.}
    \label{fig:roc}
\end{figure}

\textbf{Probing with anomaly detection.}
Following the anomaly detection setup in the original jBOT work, we use three distance metrics to measure the difference between a test jet and the bulk of the reference QCD jets as the anomaly score for detecting $W$/$Z$/$t$ jets: $k$ nearest neighbors ($k$NN), cosine similarity, and Mahalanobis distance~\citep{NEURIPS2018_abdeb6f5}.
Tab.~\ref{tab:auc-pairing} compares the ROC AUC on the combined signal across the three pairing modes, and shows that the random pairing mode has the highest performance, with the same-class mode statistically compatible.
This can be attributed to the absence of constraints: an anchor jet can pair with a jet of any class as long as the two are similar, in contrast to the class-restricted modes, which narrow the space of invariance learning.
We then optimize the metric parameters and present the results for the random-pairing model below.
Fig.~\ref{fig:score} shows the anomaly score distributions and Fig.~\ref{fig:roc} the ROC curves for the individual and combined signals, showing clear separation between QCD and signal jets.
In Tab.~\ref{tab:auc-main} we compare our pairing-based jBOT with the augmentation-based jBOT baseline.
Using the same cosine score, our method yields higher AUC for two of the three individual signal classes as well as for the combined signal, showing the potential of the pairing method to yield better separation in the embedding.

\begin{table}[!t]
  \caption{Controlled comparison of anomaly detection performance between the three pairing modes using a cosine similarity metric as anomaly score. The numbers are mean $\pm$ standard deviation over five independent pre-trainings.}
  \label{tab:auc-pairing}
  \centering
  \small
  \scalebox{0.8}{
  \begin{tabular}{llc}
    \toprule
    Mode & Pairing constraint & Combined signal AUC \\ \midrule
    Same class & $q\rightarrow q$, $g\rightarrow g$ & 0.8142 $\pm$ 0.0090 \\
    Diff class & $q\rightarrow g$, $g\rightarrow q$ & 0.7988 $\pm$ 0.0061 \\
    Random & free (no class label needed) & \textbf{0.8188 $\pm$ 0.0076} \\
    \bottomrule
  \end{tabular}
  }
\end{table}

\begin{table}[!t]
  \caption{ROC AUC comparing our pairing-based results with an augmentation-based baseline. The numbers for our models are mean $\pm$ standard deviation over three independent pre-trainings.}
  \label{tab:auc-main}
  \centering
  \small
  \scalebox{0.8}{
  \begin{tabular}{lcccc}
    \toprule
    Score & $W$ & $Z$ & $t$ & Combined \\ \midrule
    \multicolumn{2}{l}{\textit{EMD-pairing-based jBOT (ours)}}  &  & &  \\
    $k$NN & 0.7826 $\pm$ 0.0195 & 0.8002 $\pm$ 0.0105 & \textbf{0.8772 $\pm$ 0.0195} & 0.8200 $\pm$ 0.0039 \\
    Cosine & \textbf{0.8192 $\pm$ 0.0050} & 0.8154 $\pm$ 0.0083 & 0.8581 $\pm$ 0.0174 & \textbf{0.8309 $\pm$ 0.0067} \\
    Mahalanobis & 0.7659 $\pm$ 0.0089 & 0.7970 $\pm$ 0.0061 & 0.8713 $\pm$ 0.0113 & 0.8114 $\pm$ 0.0027 \\ \midrule
    \multicolumn{2}{l}{\textit{Augmentation-based jBOT (baseline)}}  &  & &  \\
    jBOT (cosine) from \citep{SciPostPhys.21.3.053} & 0.8064 $\pm$ 0.0028 & \textbf{0.8355 $\pm$ 0.0027} & 0.8388 $\pm$ 0.0028 & 0.8269 $\pm$ 0.0027 \\
    \bottomrule
  \end{tabular}
  }
\end{table}

\section{Summary}
We have presented a data-driven way to pair similar events by their energy mover's distance and shown that it is a viable substitute for handcrafted augmentations in self-supervised pre-training methods.
Pre-trained on a jet dataset with this pairing approach, the embedding shows clear separation between different jet classes including those not used in the training, and when probed for anomaly detection with simple distance-based metrics yields comparable to or better performance than a baseline under the same self-supervised pre-training method but based on augmentations.
Our results demonstrate the validity of this pairing technique for training foundation models whose embeddings are insensitive to fluctuations between physically similar events.
Future directions include applying this pairing to other types of data such as image-based detector data and event-based data to broaden its applications.

\begin{ack}

HFT and DSR are supported by the U.S. Department of Energy (DOE), Office of Science, Office of High Energy Physics subprogram on Computational High Energy Physics under Award No. DE-SC0026801.
DSR is also supported by the U.S. Department of Energy (DOE), Office of Science, Office of High Energy Physics Early Career Research program under Award No. DE-SC0025324.
This work used resources available through the National Research Platform (NRP) at the University of California, San Diego~\citep{10.1145/3708035.3736060}.
NRP has been developed, and is supported in part, by funding from National Science Foundation, from awards 1730158, 1540112, 1541349, 1826967, 2112167, 2100237, and 2120019, as well as additional funding from community partners.
Generative AI (Anthropic's Claude) was used to assist with coding and language editing.
\end{ack}


\bibliographystyle{plainnat}
\bibliography{references}








\end{document}